# Diversity of photonic differentiators based on flexible demodulation of phase signals [*]

Zheng Ao-ling (郑傲凌), Dong Jian-Ji (董建绩) [+], Lei Lei (雷蕾), Yang Ting (杨婷), and Zhang Xin-Liang (张新亮)

Wuhan National Laboratory for Optoelectronics, School of Optoelectronic Science and Engineering, Huazhong University of Science and Technology, Wuhan 430074, China

We theoretically prove a multifunctional photonic differentiation (DIFF) scheme based on phase demodulation using two cascaded linear filters. The photonic DIFF has a diversity of output forms, such as 1st order intensity DIFF, 1st order field DIFF and its inversion, 2nd order field DIFF, dependent on the relative shift between the optical carrier and the filter's resonant notches. As a proof, we also experimentally demonstrate the DIFF diversity using a phase modulator and two delay interferometers (DIs). The calculated average deviation is less than 7% for all DIFF waveforms. Our schemes show the advantages of flexible DIFF functions and forms, which may have different optical applications. For example, high order field differentiators can be used to generate complex temporal waveforms. And intensity differentiators are useful for ultra-wideband pulse generation.

## 1. Introduction

Photonic differentiation (DIFF) can offer huge bandwidth of signal processing and immunity to electromagnetic interference compared to the traditional electronic DIFF.

[*] Project supported by the National Basic Research Program of China (Grant No. 2011CB301704), the Program for New Century Excellent Talents in Ministry of Education of China (Grant No. NCET-11-0168), a Foundation for the Author of National Excellent Doctoral Dissertation of China (Grant No. 201139), and the National Natural Science Foundation of China (Grant No. 60901006, and Grant No. 11174096).

[+] Corresponding author. Email: jjdong@mail.hust.edu.cn

Thus photonic DIFF has been attracting lots of interests due to its potential wide applications in the ultrafast optical digital processing and analog processing [1-5]. The design and implementation of photonic DIFF is a primary step toward the practical optical computing circuits.

To date, photonic DIFF can be mainly divided into two categories, the optical field DIFF and the optical intensity DIFF [6]. The intensity DIFF means the output optical intensity signal is the mathematical DIFF of input optical intensity, which could be used in ultra-wideband (UWB) microwave communications [7-9] and signal encoding [10]. The intensity DIFF could be implemented by incoherent photonic processors [11], highly nonlinear fibers [7], and semiconductor optical amplifiers (SOAs) [8, 12, 13]. On the other hand, the field DIFF means the output optical field (complex signal, including both amplitude and phase) is the DIFF of input field signals, which has potential applications in ultrashort pulse generation [3, 14], a family of high order DIFF waveform generation [15-17], and pulse edge recognition [18]. To date, the field DIFFs were implemented by fiber Bragg gratings [19, 20], long-period fiber gratings [2, 21], interferometers [16, 22], SOAs [14], and silicon integrated waveguides [15, 16, 23, 24]. It is noticed that most of these aforementioned schemes were not versatile and showed a sole function. Flexible and versatile DIFF functions are desirable to meet multi-requirements in photonic signal processing. Previously, we demonstrated the field DIFF and the intensity DIFF simultaneously based on phase modulation and optical filtering [25]. However high order field DIFF is impossible in this scheme.

In this paper, to expand more DIFF functions, we theoretically deduce and experimentally prove diversity of photonic DIFF based on phase demodulation assisted by two delay interferometers (DIs). The photonic differentiator can output a diversity of DIFF waveforms, such as 1$^{st}$ order intensity DIFF, 1$^{st}$ order field DIFF and its inversion, 2$^{nd}$ order field DIFF, dependent on the relative shift between the optical carrier and the filter resonant notches. The output waveforms have more formations compared to these of Ref. [25]. Our scheme shows the advantages of flexible DIFF functions. For example, high order field differentiators can be used to generate complex temporal waveforms. And intensity differentiators are useful for UWB pulse generation.

## 2. Operation principle

The operation principle is shown in Fig. 1. A tunable laser source (TLS) emits a continuous wave (CW) with a central wavelength of $\lambda_0$. A phase modulator (PM) is driven by an electrical temporal signal $s(t)$ to modulate the CW. Two cascaded DIs with their notch wavelengths of $\lambda_{p1}$ and $\lambda_{p2}$ follow the PM. The frequency response of the DI can be approximately regarded as a linear response near the notch frequency [16]. According to the theoretical analysis of Ref. [25], the output optical power of the first DI (DI$_1$) can be expressed by

$$P_{out1} = \left(\beta \frac{\partial s(t)}{\partial t} + \omega_0 - \omega_{p1}\right)^2 \quad (1)$$

where $\omega_0$ and $\omega_{p1}$ are the optical carrier frequency and the notch angular frequency of DI$_1$, respectively, satisfying $\omega_i = 2\pi c/\lambda_i$ ($i = 0, p1$), and $\beta$ is the phase-modulated index.

With simple mathematical derivation, the output optical power of the second DI (DI$_2$) can be expressed as

$$P_{out2} = \left(\beta \frac{\partial s(t)}{\partial t} + \omega_0 - \omega_{p1}\right)^2 \left(\beta \frac{\partial s(t)}{\partial t} + \omega_0 - \omega_{p2}\right)^2 + \left(\beta \frac{\partial^2 s(t)}{\partial t^2}\right)^2 \quad (2)$$

where $\omega_{p2}$ is the central frequency of DI$_2$, satisfying $\omega_{p2} = 2\pi c/\lambda_{p2}$.

If we define that $\Delta\omega_1 = \omega_0 - \omega_{p1}$, and $\Delta\omega_2 = \omega_0 - \omega_{p2}$, then Eq. (2) can be simplified with

$$P_{out2} = \left[\left(\beta \frac{\partial s(t)}{\partial t}\right)^2 + \beta(\Delta\omega_1 + \Delta\omega_2)\frac{\partial s(t)}{\partial t} + \Delta\omega_1 \Delta\omega_2\right]^2 + \left(\beta \frac{\partial^2 s(t)}{\partial t^2}\right)^2 \quad (3)$$

Eq. (3) suggests that the output power contains lots of DIFF forms, such as 1$^{st}$ order DIFF, square of 1$^{st}$ order DIFF and 2$^{nd}$ order DIFF, and their product terms. Since Eq. (3) is a very complex equation, one may simplify it under certain specific conditions.

### A. Case 1: the first-order intensity DIFF

Eq. (3) can be simplified into four cases, which represent four different DIFF forms respectively. In the first case, assume that the wavelength notches of both DIs are located at the same sides of the wavelength of optical carrier and misaligned to the notches. To simplify Eq. (3), there are several approximation criterions, 1) the phase modulation index $\beta$ is very small, 2) the misalignment of $\Delta\omega_1$ and $\Delta\omega_2$ is very small so that the DI response meets the linear response, and 3) the output power of high order DIFF term is much lower than that of the 1$^{st}$ order DIFF term. These approximations are easily satisfied in practical experiments. Then Eq. (3) can be simplified approximately

$$P_{out2} \approx 2\beta\Delta\omega_1\Delta\omega_2(\Delta\omega_1 + \Delta\omega_2)\frac{\partial s(t)}{\partial t} + (\Delta\omega_1\Delta\omega_2)^2 \tag{4}$$

Eq. (4) suggests that the output power is the 1$^{st}$ order intensity DIFF of input signals. Then Fig. 2 calculates the output temporal waveforms of DI$_1$ and DI$_2$ when the input signal is a super-Gaussian pulse according to Eq. (3). The pulsewidth of input signal is set at 200 ps. In the simulation, we set $\lambda_{p1} - \lambda_0 = \lambda_{p2} - \lambda_0 = 0.1 nm$ in Fig. 2(a), and set $\lambda_{p1} - \lambda_0 = \lambda_{p2} - \lambda_0 = -0.1 nm$ in Fig. 2(b). One can see that the output waveform of DI$_2$ is exactly the 1$^{st}$ order intensity DIFF result. At the same time, the DIFF waveforms of Figs. 2(a) and (b) are polarity-reversed, which can be explained by Eq. (4). The polarity reversed DIFF waveforms can be used in pulse polarity modulation of UWB signals.

### B. Case 2: the first-order field DIFF

In the second case, assume that the notch of one DI is exactly aligned to the optical carrier, whereas the notch of the other DI is misaligned. Under the same approximation conditions to Case 1, Eq. (3) can be simplified with

$$P_{out2} \approx [\beta(\Delta\omega_1 + \Delta\omega_2)]^2 \left(\frac{\partial s(t)}{\partial t}\right)^2 \tag{5}$$

From Eq. (5), one can see that the output power is square of the 1$^{st}$ order DIFF waveform, indicating the output form of 1$^{st}$ order field DIFF. Fig. 3 calculates the output temporal waveforms of DI$_1$ and DI$_2$ when the input signal is a super-Gaussian pulse based on Eq. (3). In the simulation, the pulsewidth of input signal is still set at 200 ps. the wavelength misalignment of one DI is set at $\pm 0.1$nm, and the other is set at zero, as shown in Figs. 3(a)-(d) respectively. It is proved that all the output waveforms of DI$_2$ are the same, with the square of first-order DIFF waveform. This field DIFF can be used to extract the pulse edges, and suppress the signal noise with direct current (DC) in optical signal processing systems.

### C. Case 3: the first-order field DIFF in inversion

In the third case, assume that the notches of two DI are both misaligned to the optical carrier, but symmetrically distributed on the opposite sides of optical carrier. For example, we set $\lambda_{p1} - \lambda_0 = 0.1$nm, and $\lambda_{p2} - \lambda_0 = -0.1$nm. Then $\beta(\Delta\omega_1 + \Delta\omega_2)\frac{\partial s(t)}{\partial t} = 0$. With the

approximation of small phase modulation index, Eq. (3) can be simplified with

$$P_{out2} \approx 2\beta^2 \Delta\omega_1 \Delta\omega_2 \left(\frac{\partial s(t)}{\partial t}\right)^2 + (\Delta\omega_1 \Delta\omega_2)^2 \qquad (6)$$

Since $\Delta\omega_1 \Delta\omega_2 < 0$, the first term of right hand side (RHS) of Eq. (6) is opposite to the square of optical field DIFF, and the second term is a positive constant. Therefore the output waveform is a filed DIFF with an inversion. Fig. 4 shows the simulated output waveforms of $DI_1$ and $DI_2$, where $\lambda_{p1} - \lambda_0$=0.1nm, and $\lambda_{p2} - \lambda_0$=-0.1nm in Fig. 4(a), and an opposite wavelength setting is used in Fig. 4(b). For injecting a super-Gaussian pulse, the output DIFF forms show two notches in the pulse edges of input signals. This case demonstrates totally inverted waveforms compared to Case 2. So this type of DIFF may have the similar applications to that of Case 2.

### D. Case 4: the second-order field DIFF

In the fourth case, assume that the notches of two DI are both aligned exactly to the optical carrier. That is to say, we set $\lambda_{p1} - \lambda_0$=0nm, and $\lambda_{p2} - \lambda_0$=0nm. Thus $\Delta\omega_1 = \Delta\omega_2 = 0$. Then Eq. (3) can be expressed by

$$P_{out2} = \left(\beta \frac{\partial s(t)}{\partial t}\right)^4 + \beta^2 \left(\frac{\partial^2 s(t)}{\partial t^2}\right)^2 \qquad (7)$$

With the approximation of small phase modulation index, the first term of the RHS of Eq. (7) can be ignored. Hence Eq. (7) can be further simplified with

$$P_{out2} \approx \beta^2 \left(\frac{\partial^2 s(t)}{\partial t^2}\right)^2 \qquad (8)$$

From Eq. (8), one can see that the output power is square of the 2$^{nd}$ order field DIFF, indicating a second-order field DIFF. Fig. 5 shows the simulated output waveforms of $DI_1$ and $DI_2$, where $\Delta\omega_1 = \Delta\omega_2 = 0$. The output waveform of $DI_2$ accords well with the formation of the 2$^{nd}$ order field DIFF, although there is a little distortion. This distortion is caused by the term of biquadrate of 1$^{st}$ order DIFF, as indicated in Eq. (7). This kind of DIFF is a higher order DIFF, which may be useful to construct more complex temporal waveforms.

From the aforementioned analysis and mathematical inference, we have proved four kinds of DIFF formations including 1$^{st}$ order intensity DIFF, 1$^{st}$ order field DIFF and its inversion, and 2$^{nd}$ order field DIFF. These DIFF formations are dependent on the relative shift between the optical carrier and the DI resonant notches, which are summarized in

Table 1. Therefore it is very easy to alter the DIFF formation by changing the laser wavelength or the DI resonant notches.

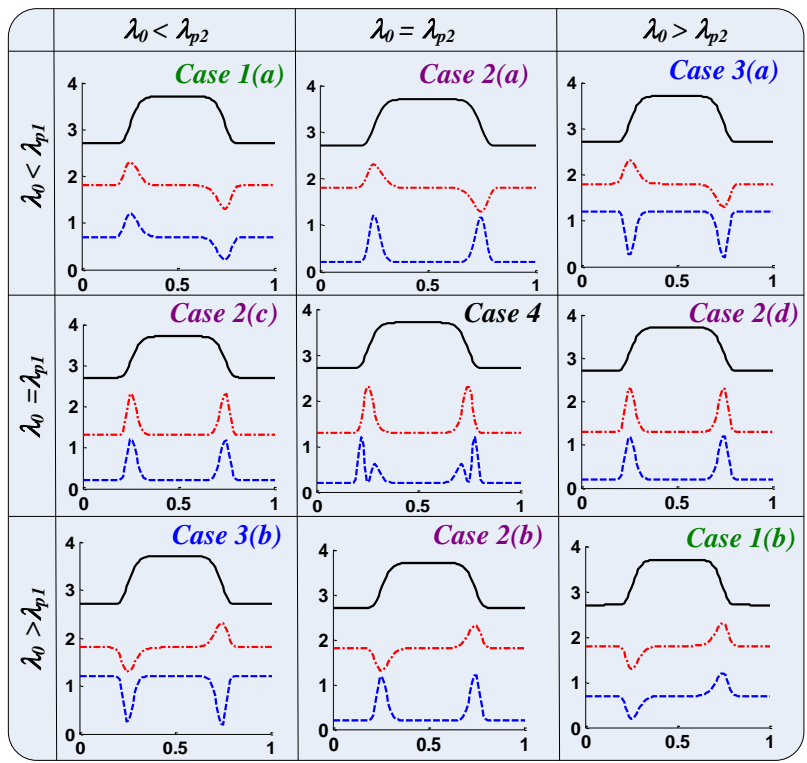

Table 1 Summary for all possible DIFF forms

## 3. Experimental demonstration and discussion

In order to verify the feasibility of our scheme, a proof-of-concept experiment is carried out, as shown in Fig. 1 as well. A CW beam is emitted from the TLS with a precisely tuning resolution of 1 pm. Then the CW beam is modulated by the PM. A polarization controller (PC) is placed before the PM to optimize the incident polarization state since the PM is polarization dependent. The PM (bandwidth: 40 GHz) is driven by an electrical super-Gaussian signal, $s(t)$, which is generated by the bit pattern generator (BPG, SHF BPG 44E). The first erbium doped fiber amplifier (EDFA) is used to boost the input optical power. Two cascaded DIs have a free spectral range (FSR) of 40 GHz and 100 GHz, respectively. The resonant frequencies of the DIs can be adjusted by the driving voltage. The output power is then optimized by the second EDFA and an attenuator (ATT), and measured by a digital communications analyzer (Agilent DCA86100C).

First the laser wavelength is set at 1563.9 nm and its output optical power is 13 dBm. The PM is driven by an electrical super-Gaussian pulse train with a repetition rate of 1.25 GHz, and the pulsewidth is about 200 ps, as shown in R1 of Fig. 6(a). Since the optical field could not be measured practically, all the DIFF results are analyzed with optical power. In order to verify Case 1, we set $\lambda_{p1} - \lambda_0 = \lambda_{p2} - \lambda_0 = 0.1 nm$. Then the output temporal waveforms of $DI_1$ and $DI_2$ are measured as shown in R2 and R4 of Fig. 6(a). The calculated waveforms are shown in R3 and R5, respectively for comparison. In the following content, R1-R5 represent the same meanings to those of Fig. 6(a) without specific statements. One can see that the measured results have good agreements with the simulated ones. This experimental result can prove the case of $1^{st}$ order intensity DIFF, corresponding to the simulations of Fig. 2(a). In the following experiment, all parameters are maintained except the resonant notches of the DIs. In order to verify the $1^{st}$ order field DIFF of Case 2, we readjust the driving voltage of the DIs so that one DI notch is aligned and the other is misaligned to the laser wavelength. In such a case, the measured output DIFF waveforms are shown in Fig. 6(b). One can see that the measured results have good agreements with the simulated ones. This experiment proves the $1^{st}$ order field DIFF, corresponding to the simulations of Fig. 3(c).

To verify Case 3, the notches of DIs are readjusted so that they are symmetrically distributed on the opposite sides of the laser wavelength. And the output DIFF waveforms are measured as shown in Fig. 7(a). One can see that the measured results have good agreements with the simulated ones. This experiment proves the $1^{st}$ order field DIFF of inversion, corresponding to Fig. 4(a). Finally, the resonant notches of DIs are adjusted so that they are exactly aligned to the laser wavelength. Fig. 8 shows the transmission spectra of $DI_1$ and $DI_2$, the input laser wavelength and the output spectrum of $DI_2$. $DI_1$ has an FSR of 40 GHz, and $DI_2$ has an FSR of 100 GHz. One can see that their resonant notches are aligned to the laser wavelength, i.e., 1563.9 nm. In such a case, the output temporal waveforms of $DI_1$ and $DI_2$ are measured, as shown in Fig. 7(b). One can see that the measured results accord with the simulated ones except some distortion. This distortion can be explained by Eq. (7). The experiment proves the case of $2^{nd}$ order field DIFF, corresponding to the simulations of Fig. 5.

To analyze the DIFF accuracy, an average deviation is defined as the mean absolute deviation of measured DIFF power from the calculated one on certain pulse period [15, 16], which is set at 1500 ps in our experiment. Then the calculated average deviations for $1^{st}$ order intensity DIFF, $1^{st}$ order field DIFF, reversed $1^{st}$ order field DIFF, and $2^{nd}$ order field DIFF are 4.27%, 6.28%, 5.76% and 6.28%. Table 2 shows all average deviations of output waveforms of $DI_1$ and $DI_2$.

Table 2  Average deviation of all the four DIFF results.

| Item | Case 1: $1^{st}$ order intensity DIFF | Case 2: $1^{st}$ order field DIFF | Case 3: reversed $1^{st}$ order field DIFF | Case 4: $2^{nd}$ order field DIFF |
|---|---|---|---|---|
| $DI_1$ output | 4.36% | 6.78% | 5.01% | 6.59% |
| $DI_2$ output | 4.27% | 6.28% | 5.76% | 6.28% |

## 4. Conclusion

We theoretically prove diversity of photonic DIFF based on phase demodulation using two cascaded linear filters. From mathematical inference and analysis, the photonic DIFF can output four DIFF formations, including $1^{st}$ order intensity DIFF, $1^{st}$ order field DIFF, reversed $1^{st}$ order field DIFF, $2^{nd}$ order field DIFF, dependent on the relative shift between the optical carrier and the filter resonant notches. As a proof, we also experimentally demonstrate the DIFF diversity using a phase modulator and two DIs. Total average deviations are less than 7% for all DIFF waveforms. Our schemes show the advantages of flexible DIFF functions, which may have different optical applications.

Figure Captions:

**Fig. 1.** Experimental setup for the proposed multifunctional differentiators

**Fig. 2.** Fig.2 Simulated waveforms for 1$^{st}$ order intensity DIFF; (a) $\lambda_{p1} - \lambda_0 = \lambda_{p2} - \lambda_0 = 0.1nm$; (b) $\lambda_{p1} - \lambda_0 = \lambda_{p2} - \lambda_0 = -0.1nm$.

**Fig. 3.** Simulated waveforms for 1$^{st}$ order field DIFF, (a) $\lambda_{p1} - \lambda_0 = 0.1$nm, $\lambda_{p2} - \lambda_0 = 0$nm, (b) $\lambda_{p1} - \lambda_0 = -0.1$nm, $\lambda_{p2} - \lambda_0 = 0$nm, (c) $\lambda_{p1} - \lambda_0 = 0$nm, $\lambda_{p2} - \lambda_0 = 0.1$nm, (d) $\lambda_{p1} - \lambda_0 = 0$nm, $\lambda_{p2} - \lambda_0 = -0.1$nm.

**Fig. 4.** Simulated waveforms for 1$^{st}$ order field DIFF with an inversion, (a) $\lambda_{p1} - \lambda_0 = 0.1nm$, $\lambda_{p2} - \lambda_0 = -0.1nm$, (b) $\lambda_{p1} - \lambda_0 = -0.1nm$, $\lambda_{p2} - \lambda_0 = 0.1nm$.

**Fig. 5.** Simulated waveforms for 2$^{nd}$ order field DIFF, $\lambda_{p1} - \lambda_0 = \lambda_{p2} - \lambda_0 = 0$.

**Fig. 6.** (a) Experimental results to prove Case 1, (b) Experimental results to prove Case 2

**Fig. 7.** (a) Experimental results to prove Case 3, (b) Experimental results to prove Case 4

**Fig. 8.** Input and output spectra of 2$^{nd}$ order field DIFF.

**Figures:**

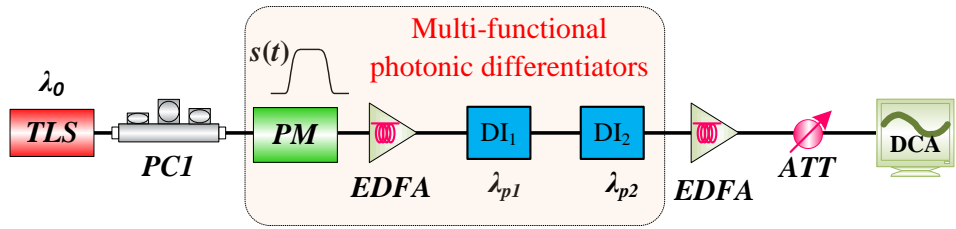

**Fig. 1**

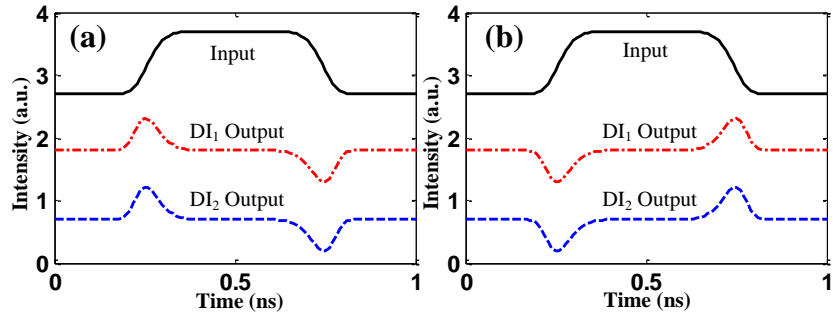

**Fig. 2**

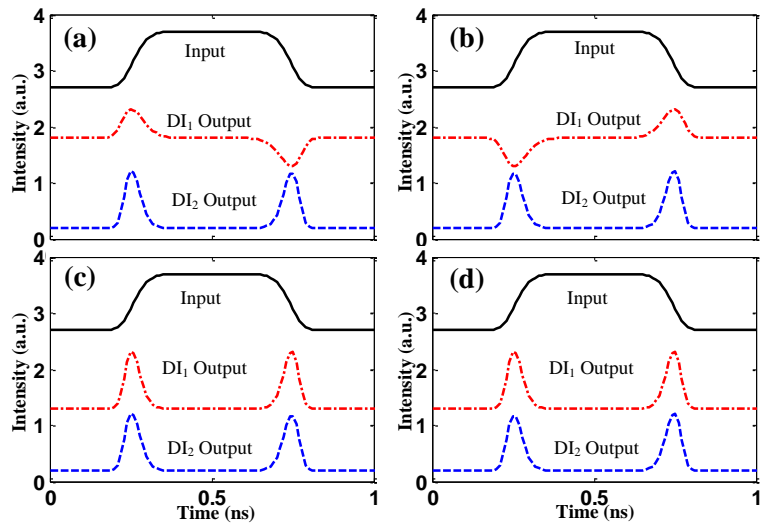

**Fig.3**

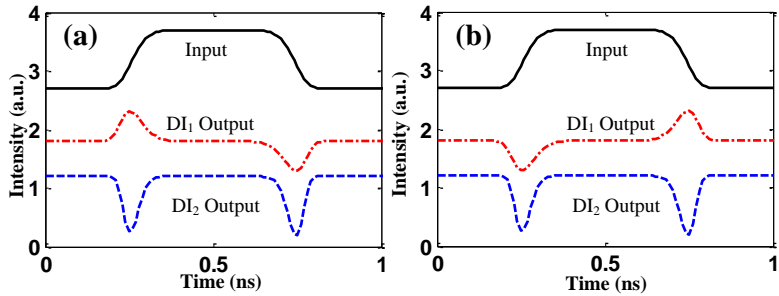

**Fig.4**

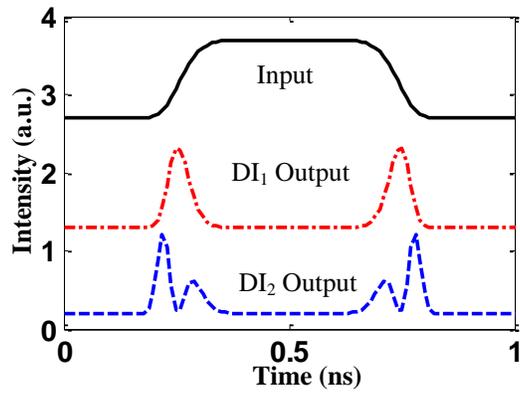

Fig.5

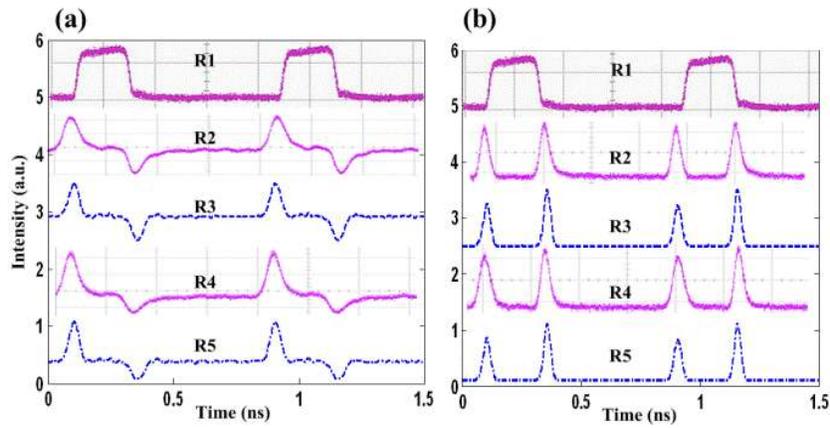

Fig.6

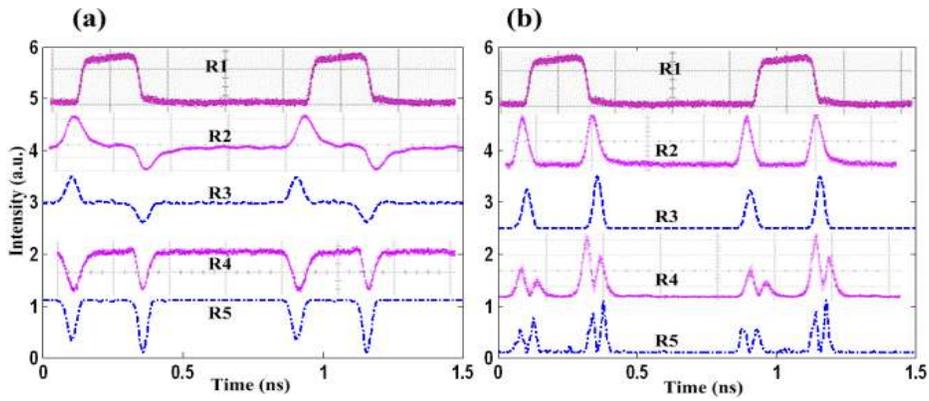

Fig. 7

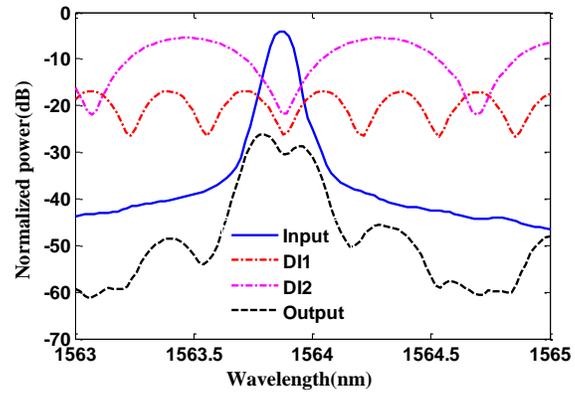

**Fig. 8**